%% file: main20261119.tex
\documentclass[aps,prd,nofootinbib,twocolumn,superscriptaddress,letterpaper,preprintnumbers]{revtex4-1}
\usepackage{graphicx,dcolumn,bm,float}
\usepackage{anyfontsize}
\usepackage{color}
\usepackage{xcolor}
\usepackage{amsmath, amssymb,braket, mathtools, soul}
\usepackage{dsfont}
\usepackage[dvipsnames]{xcolor}
\usepackage[svgnames]{xcolor}
\usepackage{physics}

\usepackage{color}
\usepackage{float}
\usepackage{graphicx}
\usepackage[caption=false]{subfig}
\usepackage{array,booktabs,tabularx}
\usepackage{natbib}
\usepackage{comment}

\usepackage[colorlinks=true
,urlcolor=blue
,anchorcolor=blue
,citecolor=blue
,filecolor=blue
,linkcolor=red
,menucolor=blue
,hyperfootnotes=false,
,linktocpage=true
,pdfproducer=medialab
,pdfa=true
]{hyperref}

\usepackage{appendix}

\usepackage{tikz}
\usetikzlibrary{calc,decorations.pathmorphing,arrows.meta}

\input{tikz_images}

\input{trispectrum_momenta}

\newcommand{\beq}{\begin{equation}}
\newcommand{\eeq}{\end{equation}}
\newcommand{\bea}{\begin{eqnarray}}
\newcommand{\eea}{\end{eqnarray}}
\newcommand{\bef}{\begin{figure}}
\newcommand{\eef}{\end{figure}}

\newcommand{\mpl}{M_{\rm Pl}}
\newcommand{\Lamsp}{\Lambda_{\rm sp}}

\newcommand{\A}{\mathcal{A}}
\newcommand{\Aodd}{\mathcal{A}^{\rm odd}}
\newcommand{\AoddD}{\mathcal{A}^{\rm odd}_{\rm detect}}
\newcommand{\Pic}{\Pi_{\rm c}}
\newcommand{\PicS}{\Pi_{\rm c}^{\rm str}}

\begin{document}

\title{UV Bounds on Inflationary Parity Violation}

\author{Alexander Cassem}
\email{alexander.cassem@tufts.edu}
\author{Mark P.~Hertzberg}
\email{mark.hertzberg@tufts.edu}
\affiliation{Institute of Cosmology, Department of Physics and Astronomy, Tufts University, Medford, MA 02155, USA
\looseness=-1}

\begin{abstract}
A question of fundamental interest is whether parity is violated on large scales in the Universe. This possibility has been motivated by recent analyses of the four-point matter correlation function. It has also been suggested that detectable parity violation could be generated during inflation by the dynamical Chern-Simons operator, a leading effective field theory correction to general relativity. In this work, we examine this possibility in realistic UV completions, including fermionic and string-theoretic examples. We show that existing bounds on the tensor-scalar-scalar (TSS) bispectrum strongly constrain the possibility of generating a detectable parity-odd trispectrum. For tensor-to-scalar ratios near current observational limits, detectability is incompatible with, or pushes against, the regime of validity of these UV completions. A detectable signal can remain possible for a sufficiently small tensor-to-scalar ratio, but only at the cost of an extremely large enhancement of the TSS vertex. Thus, while a future search for large scale parity violation is worth pursuing, generating a detectable signal within a consistent effective theory is highly nontrivial.
\end{abstract}

\maketitle


\section{Introduction}

The leading theory of the early universe is cosmic inflation, a phase of approximate exponential expansion, which helps to explain the large scale homogeneity, isotropy, and flatness of the universe \cite{Starobinsky:1980te, Guth:1980zm,Linde:1981mu,Albrecht:1982wi,Linde:1983gd,Baumann:2009ds}. Quantum fluctuations during inflation leads to small deviations away from homogeneity and isotropy \cite{Mukhanov:1981xt,Hawking:1982cz,Starobinsky:1982ee,Guth:1982ec,PhysRevD.28.679}. In the simplest models, these quantum fluctuations are approximately scale invariant and Gaussian. 
The amplitude ($A_s$) and tilt ($n_s$) are well measured \cite{Planck:2018vyg,Planck:2018jri}, which in the simplest slow-roll models, they constrain
only two combinations of parameters in the inflationary Lagrangian. 
This leaves many models compatible with observations. It is therefore of central importance to observe deviations from the minimal predictions, in order to gain an understanding of the details of the inflationary dynamics.  

An important possibility is that there could be non-Gaussianity in the fluctuations which is encoded in higher point correlation functions, including the 3-point bispectrum and/or the 4-point trispectrum functions \cite{Maldacena:2002vr,Senatore:2009gt,Smith:2015uia}. Interestingly, if there is parity violation in the large scale structure, then assuming statistical homogeneity and isotropy, this can first only appear in the 4-point trispectrum. Some evidence of this possibility has emerged in recent years, although it is unconvincing at this stage \cite{Coulton:2023oug,Diego-Palazuelos:2022cnh,Hou:2022wfj,Minami:2020odp,Philcox:2022hkh,Krolewski:2024paz}. 
It is interesting to turn to the future prospects of a clear detection of observing parity violation in the 4-point galaxy trispectrum as large scale structure observations have more modes to probe compared to the CMB \cite{Sailer:2021yzm,Cahn:2021ltp}. 

In this work, we will carefully analyze future prospects of detecting large scale parity violation.  
We find that it is rather difficult from multiple point of views. In particular, we will show that existing bounds on the scalar-scalar-tensor bispectrum restrict the parity violating part of the scalar trispectrum to be below reasonable future expectations. 

A basic way to try to generate significant parity violation is to use the parity violating operator $\alpha\, \phi\, ^*\!R\,R$, where $\phi$ is the inflaton and $^*\!R\,R$ is the Pontryagin density, and when coupled together gives the dynamical Chern-Simons (dCS) operator. This is a type of leading order correction to general relativity. It can be generated in the effective theory by integrating out chiral fermions,
and it appears in type I and heterotic string compactifications \cite{Green:1984sg,Svrcek:2006yi,Alvarez-Gaume:2022aak} and in some type IIB effective actions through D-brane  coupling  \cite{Stefanski:1998he,Weissenbacher:2020cyf}.

In a rolling inflaton background, this dCS operator leads to a correction to the graviton propagator.
Thus, it can alter the 4-point correlation function from graviton exchange between a pair of graviton-inflaton-inflaton vertices. Since the operator violates parity, it makes the graviton propagator chiral which enhances one of the graviton helicities over the other \cite{Satoh:2008ck,Bartolo:2017szm,Alexander:2004wk}. In turn, this generates a 4-point correlation function that contains a parity-odd piece \cite{Creque-Sarbinowski:2023wmb}. At first sight, this might indicate one can generate an arbitrarily large parity-odd signal by making the dCS coupling, $\alpha$, arbitrarily large. However, this has a basic problem: (i) By making $\alpha$ large, one is also generating a large amount of tensor modes, which must remain below the existing bound. This was accounted for in the prior literature by noting that the parity-odd term cannot exceed that of the parity-even term. The parity odd term is $2\Theta_\text{odd} =P_+-P_-$, while the parity even term is $2\Theta_\text{even} = P_++P_-$. The ratio $\Pic= \Theta_\text{odd}/\Theta_\text{even} = (P_+-P_-)/(P_++P_-)$ by definition must obey $|\Pic|\leq1$. While this is somewhat restrictive, it still allows for a relatively large parity-odd trispectrum. In this work, we identify a second basic problem: (ii) constraints from the UV demand that $|\Pic|\ll1$ in order to have a consistent effective field theory (EFT). We use two illustrative classes of UV completions: chiral fermions and string theory compactifications. In each case we show that a consistent EFT requires a rather small dCS coefficient. This pushes the effective theory to an extreme corner of parameter space in order for detectability. We further identify that an existing bound on the (even) tensor-scalar-scalar (TSS) bispectrum constrains the structure significantly further.

Our paper is organized as follows:
In Section \ref{LSPV} we review the requirements for a possible future detection of dCS.
In Section \ref{TrispectrumBound} we describe the consequences from an existing bound on the TSS bispectrum.
In Section \ref{Fermionic} we study a fermionic UV completion.
In Section \ref{Stringy} we study a string-theoretic UV completions.
Finally, in Section \ref{Discussion} we discuss our results, and possible ways dCS can still play an important role during inflation.

\section{Large Scale Parity Violation}\label{LSPV}

To begin, let us consider an inflaton scalar $\phi$ coupled to gravity. In addition we will include the leading parity violating operator $\phi\,^*\!R\,R$. The action is then \cite{Alexander:2009tp}
\bea
S=\int d^4x\sqrt{-g}
\Bigg{[}M_{\rm Pl}^2R-{1\over2}(\partial\phi)^2-V(\phi)\nonumber\\
\,\,\,\,\,\,\,+ {\alpha M_{\rm Pl}\over4}\phi\,^*\!R\,R\Bigg{]},
\eea
where $M_{\rm Pl}=1/\sqrt{16\pi G}$ is the  Planck mass and $\alpha$ is the dCS parameter, which has units of length-squared.

If there is parity violation, such as the above dCS term, it is important to check how this alters the density correlation functions. Statistical homogeneity and isotropy ensure that the 2-point and 3-point correlation functions are parity invariant. The first correlation function that permits parity violation is the 4-point function, ie., the trispectrum. 

Let $\zeta({\bf k})$ be the fractional density fluctuation (in $k$-space). The (connected) trispectrum is a function of 4 wave vectors ${\bf k}_1,\,{\bf k}_2,{\bf k}_3,\,{\bf k}_4$ \cite{Seery:2008ax,Cabass:2022rhr}
as
\beq
\langle\zeta_1\,\zeta_2\,\zeta_3\,\zeta_4\rangle
=(2\pi)^3\delta^3\!\left(\sum_{i=1}^4{\bf k}_i\right)\,T({\bf k}_1,{\bf k}_2,{\bf k}_3,{\bf k}_4),
\eeq
where $T$ is the amplitude.
Consider a graviton-scalar-scalar vertex. Working to tree-level, a 4-point trispectrum can be constructed for the inflaton that exchanges a graviton between these vertices.
We can write the parity even and parity odd parts of this trispectrum as
$T=T^{\rm even}+T^{\rm odd}$ with \cite{Creque-Sarbinowski:2023wmb,Cabass:2022rhr}
\bea
T^{\rm even}&=&\A^{\rm even}\,S^{\rm even}\label{Teven},\,\,\,\,\,\mbox{with}\,\,\,\,\A^{\rm even}=r\,F,\\
T^{\rm odd}&=&i\,\A^{\rm odd}\,S^{\rm odd}
\label{Todd},\,\,\,\,\,\mbox{with}\,\,\,\,\A^{\rm odd}=\Pi_{\rm c}\,r\,F.
\eea
In these expressions we have factored out  the tensor-to-scalar ratio $r$, as they are produced by graviton exchange, and we have introduced a pair of shape functions $S^{\rm even},\,S^{\rm odd}$. 
Let us consider the ``collapsed limit" in which ${\bf q}={\bf k}_1+{\bf k}_2=-({\bf k}_3+{\bf k}_4)$ obeys 
\beq
q\ll \{k_1,k_2,k_3,k_4\}
\eeq
or visually
\begin{equation}
\vcenter{\hbox{%
  \TrispectrumMomenta[
    scale=0.55,
    q=0.65,k3=0.90,
    theta1=1.15,theta3=1.40,chi=2.40,
    ratios=false
  ]%
}}
\quad
\xrightarrow{\;q/k_1\to0\;}
\quad
\vcenter{\hbox{%
  \TrispectrumMomenta[
    scale=0.55,
    q=0.08,k3=0.90,
    theta1=1.15,theta3=1.40,chi=2.40,
    ratios=false
  ]%
}}\label{eq: collapsed limit}\nonumber
\end{equation}
This is the limit in which the exchanged graviton has very small momenta. One can show that in this collapsed limit, the shape functions take the form
\bea
S^{\rm even}&=&\cos(2\chi)\,\mathcal{S},\\
S^{\rm odd}&=&\sin(2\chi)\,\mathcal{S},
\eea
where
\beq
\mathcal{S}=\frac{9}{16}\,
\sin^2\theta_1\sin^2\theta_3\,
P_\zeta(q)P_\zeta(k_1)P_\zeta(k_3).
\eeq
Here, $P_\zeta$ is the scalar power spectrum, 
$\theta_1$ and $\theta_3$ are the angles between the hard
momenta $\mathbf{k}_1,\mathbf{k}_3$, and the soft exchanged momentum $\mathbf q$
namely $\mathbf{k}_i\cdot\mathbf{q} = k_i\,q\,\cos\theta_i$, and $\chi$ is the relative azimuthal angle of the two
hard-momentum pairs about $\mathbf q$. In the above, we see that under a parity transformation $\chi\to-\chi$, we have $S^{\rm even}\to S^{\rm even}$ (from  $\cos(2\chi)$) and $S^{\rm odd}\to-S^{\rm odd}$ (from  $\sin(2\chi)$), as expected.

Note that in Eqs.~\eqref{Teven} and \eqref{Todd}, we have introduced the factor $F$. For canonical models, $F=1$, but one can imagine the possibility of $F\gg1$ in modified models in which the graviton-scalar-scalar vertex is enhanced. We shall return to this point later.

Finally, in Eq.~(\ref{Todd}), we have introduced the chirality factor $\Pi_{\rm c}$, which is given in terms of the power of the 2 helicities of the graviton as
\beq
\Pi_{\rm c}=\frac{\Theta_\text{odd}}{\Theta_\text{even}}={P_+-P_-\over P_++P_-},\label{eq: circular polarization}
\eeq
where $P_+$ is the power spectrum with the $+$ helicity and $P_-$ is the power spectrum in the $-$ helicity. We note that $|\Pi_{\rm c}|\leq1$, and so this construction forbids the parity-odd piece from being larger than the parity-even piece since the exchanged graviton contributes to both. As mentioned prior, the even piece is $2\Theta_\text{even} = P_+ + P_-$ while the odd piece is $2\Theta_\text{odd} = P_+ - P_-$. The even piece can be related to the tensor-to-scalar ratio by $r = (P_+ + P_-)/P_\zeta$.

Now, if we have no dCS term (i.e., we set $\alpha=0$) then we can still have an even trispectrum from ordinary graviton exchange, but a vanishing odd trispectrum.
For a nonzero dCS term, one can expand around the inflaton background and identify that the dCS term modifies the graviton propagator. This alters the  trispectrum as it arises from graviton exchange, which will alter both the even and odd part. The correction to the even part is small at weak coupling due to a cancellation, but the odd part is generated at linear order in the coupling. The chirality factor at weak coupling (we shall justify the weak coupling assumption shortly) is given by \cite{Alexander:2004wk,Satoh:2008ck}
\beq
    \Pi_{\rm c}
    =
    \frac{\pi}{2}\alpha\, H^2\sqrt{\epsilon}.
    \label{eq:Pi-alpha}
\end{equation}
Here, $H$ is the Hubble parameter during inflation and $\epsilon=-\dot{H}/H^2$ is the first slow-roll parameter. It is intuitive
that the result is proportional to $\sqrt{\epsilon}$, since (for a canonical scalar) one has $\sqrt{\epsilon}=|\dot{\phi}|/(2\,H\,M_{\rm Pl})$ and the dCS term only affects the equation of motion when the field is rolling ($\dot\phi\neq0$). Otherwise, the dCS term is a total derivative and can be ignored when $\dot{\phi}\approx0$.

\subsection{Prospects for Detectability}

For the parity-odd graviton-exchange trispectrum, the signal-to-noise ratio
may be written  as \cite{Smith:2015uia}
\begin{equation}
    \left(\frac{S}{N}\right)^2
    =
    \frac{(\A^{\rm odd})^2}{24}\,
    N_{\rm modes}\,
    {\cal N}_{\rm GE},
    \label{eq:SN-general}
\end{equation}
where $N_{\rm modes}$ is the number of measured Fourier modes and
${\cal N}_{\rm GE}$ is a dimensionless integral over the
graviton-exchange shape.  In the collapsed limit, \cite{Creque-Sarbinowski:2023wmb} finds
\begin{equation}
    {\cal N}_{\rm GE}
    \simeq
    \frac{9}{200}\,
    A_s^2
    \left(\frac{k_{\rm max}}{k_{\rm min}}\right)^3 ,
    \label{eq:NGE-collapsed}
\end{equation}
where
\begin{equation}
    A_s\simeq 2.1\times10^{-9}
\end{equation}
is the measured amplitude of the primordial scalar power spectrum. Here, $k_{\rm max}$ is the maximum wavenumber of a survey and $k_{\rm min}$ is the minimum. 

Re-arranging to solve for the amplitude of the parity-odd trispectrum gives
\begin{equation}
    |\A^{\rm odd}|
    \approx 23
    \frac{S/N}{A_s\sqrt{N_{\rm modes}}}
    \left(\frac{k_{\rm min}}{k_{\rm max}}\right)^{\!3/2}.
    \label{eq:Pi-r-F-parametric}
\end{equation}
Now the number of modes is closely tied to the ratio of the maximum and minimum wavenumbers by
\beq
N_{\rm modes}\sim\left(k_{\rm max}\over k_{\rm min}\right)^{\!3},
\eeq
but we shall keep them separate for completeness. 
If we would like a detection with signal-to-noise ratio of at least $S/N=3$, then we require
\beq
|\A^{\rm odd}|\geq \AoddD,\label{eq: A odd detect}
\eeq
where the minimal amplitude for a detection is
 \beq
 \AoddD
 \approx 3.3\times 10^4
 \left(10^3\over\sqrt{N_{\rm modes}}\right)
 \left(100\over k_{\rm max}/k_{\rm min}\right)^{\!3/2},
\label{Adetect}\eeq
where we have used a number of modes of $10^6$ corresponding to the CMB, $k_{\rm min}\sim 0.003\,h$/Mpc, and $k_{\rm max}\sim 0.3\,h$/Mpc as reasonable targets. 
So, a model in which the parity-odd trispectrum amplitude of $|\A^{\rm odd}|=|\Pi_c|\,r\,F$
is to be potentially measurable in future surveys, must obey $|\mathcal{A}_\text{odd}|\gtrsim 3.3\times 10^4$. 

One may wonder if this is possible within a theory. Firstly, we know that the tensor-to-scalar ratio $r$ is bounded by \cite{Tristram:2021tvh}
\beq
r\leq 0.032\,\,\,\,(\mbox{at}\,\,95\%)
\eeq
from combining Planck PR4, BICEP/Keck 2018, SDSS/eBOSS BAO, and Planck CMB lensing. Now, recalling that $|\Pi_{\rm c}|\leq 1$, we can rewrite the parity-odd trispectrum amplitude as
\beq
\A^{\rm odd}\approx 3.3\times 10^4\left(\Pi_{\rm c}\over 1\right)
\left(r\over 0.032\right)\left(F\over10^6\right).
\eeq
This shows that one needs a very large enhancement, $F\gtrsim 10^6$, in order to achieve a future detection. 

If one takes the point of view that $F$ is a model dependent free parameter, then one can simply imagine that this huge enhancement factor is achievable. 
For example, in Ref.~\cite{Creque-Sarbinowski:2023wmb} it is suggested that a secondary field could generate an enhancement factor of $F\sim N_e^4$, where $N_e\sim 60$ is the number of e-folds of inflation. If this is achieved, then one indeed has $F\sim 60^4\approx 1.3\times 10^7$, which appears to be sufficiently large. 

It is worth investigating where this factor of $F$ could come. Consider another scalar species $\chi$ that contributes to the isocurvature perturbation with an effective mass related to the conformal index $\Delta = 3/2 - \sqrt{9/4 - m_\text{eff}^2/H^2}$. The superhorizon solution for $\chi$ is approximately $f_\chi\propto e^{-\Delta N_e}$ up to some $k$-dependent function. 
If the dimensionless isocurvature perturbation is $\mathcal{S} = Hf_\chi/\dot{\phi}$, then density perturbations are related to isocurvature roughly as $d\zeta/dN_e\propto \mathcal{S}$ \cite{Gordon:2000hv}. For slow-roll inflation, we can integrate over $dN$ to find that the final isocurvature mode is proportional to $N_e$. For the trispectrum, we would then get a contribution of $N_e$ per leg, or in other words, $F\sim N_e^4$.

However, in the next sections we shall show that in any reasonable UV completion, combined with known bounds on the parity-even trispectrum, the above required value of $|\A^{\rm odd}|\gtrsim 3.3\times 10^4$  is very difficult to achieve.

\section{Existing Bound on Parity-Even Trispectrum}\label{TrispectrumBound}

In the above, we reviewed the required level of enhancement needed for the parity-odd trispectrum to be measurable in upcoming surveys. It is important to compare this to the parity-even part, and of course, we must be consistent with {\em existing} bounds on the parity-even trispectrum \cite{Philcox:2025bvj,Philcox:2025lrr,Philcox:2025wts}.

To make progress on this, let us note that in the collapsed limit, the trispectrum factorizes into a product of the tensor-scalar-scalar bispectrum $B_{\gamma\zeta\zeta}$ as
\beq
T
\simeq
\sum_{\lambda=\pm}
\frac{
B_{\gamma\zeta\zeta}^{(12)}
B_{\gamma\zeta\zeta}^{(34)}
}{
P_{\gamma} },
\eeq
where the sum is over the 2 helicities of the graviton.
Now the tensor-scalar-scalar bispectrum is given in terms of the corresponding amplitude parameter $f_{\gamma\zeta\zeta}^{\rm NL}$. It is defined with respect to the leading value in the gravitational EFT, namely GR,
so that 
\beq
{B_{\gamma\zeta\zeta}\over B_{\gamma\zeta\zeta,\rm GR}}={f_{\gamma\zeta\zeta}^{\rm NL}\over f_{\gamma\zeta\zeta,\rm GR}^{\rm NL}}.
\eeq
The general relativity value is adopted to have the normalization
\beq
f_{\gamma\zeta\zeta,\rm GR}^{\rm NL}=
-{5\over 32\sqrt{2}}r.
\eeq
Taking the even part of the trispectrum, we then get \cite{Philcox:2024wqx}
\beq
T^{\rm even}=
\left(B_{\gamma\zeta\zeta}\over B_{\gamma\zeta\zeta,\rm GR}\right)^2
T_{\rm GR}
=\left({32\sqrt{2}\over 5r}f_{\gamma\zeta\zeta}^{\rm NL}\right)^2\,r\,\mathcal{S}^{\rm even}.
\eeq
Then, using $T^{\rm even}=\A^{\rm even}\,S^{\rm even}=r\,F\,S^{\rm even}$, we obtain
\beq
r\,F={2048\over 25r}
\left(f_{\gamma\zeta\zeta}^{\rm NL}\right)^2
\label{rFfNL}
\eeq
(which is just the statement that $F$ is the enhancement factor $B_{\gamma\zeta\zeta}=\sqrt{F}\,B_{\gamma\zeta\zeta,\rm GR}$). 

Current data gives an observational  bound on the tensor-scalar-scalar bispectrum.
The Planck PR4 analysis gives a bound in the squeezed limit of (which relates to the trispectrum collapsed limit)
\begin{align}
f_{\gamma\zeta\zeta}^{\rm NL}
&=
11\pm11,
\qquad {\rm SEVEM},
\\
f_{\gamma\zeta\zeta}^{\rm NL}
&=
13\pm10,
\qquad {\rm SMICA}.
\end{align}
So, there is currently no statistically significant evidence for a primordial TSS
bispectrum.  We shall use the latter bound which provides a conservative bound of
\begin{equation}
\left|f_{\gamma\zeta\zeta}^{\rm NL}\right|
\leq 33,\,\,\,\,(\mbox{at}\,\,95\%).
\end{equation}
Multiplying Eq.~(\ref{rFfNL}) by $\Pi_{\rm c}$ gives an expression for the odd amplitude trispectrum in terms of the tensor-scalar-scalar bispectrum of
\beq
\A^{\rm odd}\approx
2.8\times 10^6
\left(\Pi_{\rm c}\over 1\right)
\left(0.032\over r\right)\left(f_{\gamma\zeta\zeta}^{\rm NL}\over 33\right)^2
\label{AoddrF}\eeq

We note that if we are at this boundary of $\A^{\rm odd}\sim 2.8\times 10^6$, by taking $|\Pic|\sim1$, then $|\Aodd|$ is significantly higher than the required value for detectability as given in \eqref{Adetect}, $\AoddD\sim3.3\times 10^4$.
However, this hinges upon pushing the chirality between these tensor modes $|\Pic|$ towards its maximum allowed value. In the next section we shall examine if this is actually plausible in realistic UV completions.

\section{Fermionic UV Completion}\label{Fermionic}

A weakly coupled chiral fermionic UV completion coupled to a scalar that is all minimally coupled to gravity when integrated out gives the dCS operator. This interaction was studied in Ref.~\cite{Cassem:2026aib} and \cite{Alexander:2004us,Alexander:2022cow,Alexander:2024vav,Toms:2018wpy}. The corresponding action is
 \bea
    S &=& \int d^4x\sqrt{-g}\Bigg{[}\mpl^2\,R-{1\over2}(\partial\phi)^2 \nonumber\\
    &&+\sum_n\overline\psi_n(i\,\gamma^a\nabla_a - m_n +i\,g_n\,\phi\,\gamma_5)\psi_n\Bigg{]}
    \label{eq: uv completion action of dCS}
\eea
in which the scalar $\phi$ carries a pseudo-Yukawa coupling to the fermions.
This can also be understood as the leading order part of a theory with a spontaneously broken $U(1)$ symmetry. This can be seen by introducing a complex scalar field $\Phi$ with U(1) symmetry $\Phi\rightarrow e^{i\theta_0}\Phi$ and $\psi_n\rightarrow e^{i\theta_0\gamma_5/2}\psi_n$. If we choose a symmetry breaking potential $V(|\Phi|)\propto (|\Phi|^2-f^2)^2$, and expand around the vacuum $\Phi = fe^{i\phi/f}$, then we recover \eqref{eq: uv completion action of dCS}.

By integrating out these massive fermions, and setting their masses $m_n$ and couplings $g_n$ to all be the same for simplicity ($m_\psi\equiv m_1=m_2=\ldots, \,\,\,g\equiv g_1=g_2=\ldots$), one generates a dCS term with parameter
\beq
|\alpha| \mpl
=
\frac{gN}{192\pi^2m_\psi}.\label{fermionalpha}
\end{equation}
(with $N=2N_D$ the number of Weyl fermions).
Perturbative unitarity demands that the Yukawa coupling obeys 
\begin{equation}
\lambda
\equiv
\frac{N g^2}{(4\pi)^2}
<1.
\end{equation}
If there are many species, then the fundamental gravitation cut off is reduced to the ``species scale'' \cite{Dvali:2007hz}
\begin{equation}
\Lamsp
=
b\,\frac{4\pi \mpl}{\sqrt{N}},
\end{equation}
where \(b\) is an order-one coefficient. To ensure perturbativity, one expects $b\sim 1$, and from black hole arguments, one expects $b\sim 1/(4\pi)$. Using these parameters, the dCS coefficient can be rewritten as
\begin{equation}
|\alpha|
=\frac{b\sqrt{\lambda}}
     {12\,m_\psi\Lambda_{\rm sp}}.
\label{eq:alpha-bound}
\end{equation}
Since $b\lesssim 1$ and $\lambda<1$, this provides a bound on $\alpha$ in terms of the species bound $\Lambda_{\rm sp}$. 
Inserting this bound into the expression for the chirality parameter in Eq.~(\ref{eq:Pi-alpha}), gives
\beq
|\Pic|\approx
5.9\times 10^{-3}
\left(b\over 1\right)\left(\sqrt{\lambda}\over1\right)
\left(H\over m_\psi\right)\left(H\over\Lamsp\right)
\sqrt{16\epsilon\over 0.032}
\label{Pidet}\eeq
In the final term we have normalized $16\epsilon$ to the current bound on the tensor-to-scalar ratio of 0.032 \cite{Tristram:2021tvh}. In canonical models of inflation, one has $r=16\epsilon$. 
Note that consistency requires each of the terms in brackets to be at most 1. In fact in the weakly coupled regime one should have the parametric hierarchy $H\ll m_\psi$ and $H\ll\Lamsp$ to rigorously trust the effective theory. This implies $|\Pic|\ll1$, and shows that the $|\Pic|\sim 1$ assumption of \cite{Creque-Sarbinowski:2023wmb} cannot be realized in this framework. This also self-consistently justifies the usage of the weak coupling approximation for $\Pic$.

Inserting \eqref{Pidet} into Eq.~(\ref{AoddrF}) gives
\beq
|\Aodd| \approx 1.6\times 10^4
\,X\,Y\,Z
\eeq
where we have defined
\bea
&&X\equiv\left(0.032\over r\right)\sqrt{16\epsilon\over 0.032}\label{Xdef}\\
&&Y\equiv\left(b\over 1\right)\left(\sqrt{\lambda}\over1\right)
\left(H\over m_\psi\right)\left(H\over\Lamsp\right)\\
&&Z\equiv
\left(f_{\gamma\zeta\zeta}^{\rm NL}\over 33\right)^{\!2}\label{Zdef}
\,\,\,\,\,\,
\eea
For $r$ near its upper bound of $0.032$ (and comparable value of $16\epsilon$), $X\sim1$. This implies $|\Aodd|\ll 1.6\times10^4$, as all the factors in $Y$ and $Z$ should be less than 1. Then by \eqref{eq: A odd detect} this gives
\beq
|\Aodd|\ll\AoddD\,\,\,\,\,\,(\mbox{for}\,\, X\sim 1)
\eeq
i.e., much less than the value needed for detectability in Eq.~(\ref{Adetect}).

For $r$ much less than its upper bound, $|\mathcal{A}^\text{odd}|$ can be larger. For concreteness, let us consider $r\approx 16\epsilon$ (the canonical relation), then $X$ can be large and the situation is more promising. We then have $X = \sqrt{0.032/r}$,
so $r\ll 0.032$ gives $X\gg1$ and allows for a larger $\Aodd$. While this seems promising, we should pay attention to the corresponding enhancement factor $F$ that this corresponds to.

Taking $\AoddD\approx 3.3\times 10^4$, we obtain the condition on $X$ for detectability $|\Aodd|\geq\AoddD$ becomes  
\beq
X\,Y\,Z\geq 2.
\eeq
Rewriting $F$ using Eq.~(\ref{rFfNL}) in terms of the $X,Y,Z$ variables, this gives
\beq
F\approx 8.6\times 10^7\,X^4\,Z.
\label{FXZform}\eeq
By imposing the detectability condition in the form $XYZ\geq2$, we obtain
\beq
F\geq1.4 \times 10^9\,Y^{-4}Z^{-3}.
\eeq
Since $Z\leq1$ by the existing bound on tensor-scalar-scalar bispectrum and $Y\ll1$ for a weakly coupled UV completion, we have the requirement 
\beq
F\gg 10^9\,\,\,\,(\mbox{chiral fermions in UV})
\eeq
for detectability. If we consider the case of $F\sim N_e^4$, with $N_e\sim60$ giving $F\simeq 1.3\times 10^7$, then this condition is not satisfied. Such extreme enhancement factors seem difficult to achieve.

\section{String-Theoretic UV Completion}\label{Stringy}

A qualitatively distinct UV completion of dynamical Chern--Simons gravity
is provided by string theory. 
It is well known that the Green-Schwarz mechanism requires a dCS-like operator to appear in the type 1 and heterotic string theories \cite{Green:1984sg,Gross:1986mw,Polchinski:1998rr}. 

\subsection{Heterotic}
In heterotic string theory, compactification
of the ten-dimensional effective action on a $T^6$ (toroidal) geometry, generates a 4-dimensional coupling of the axion (after dualizing the Kalb-Ramond 2-form $B$-field)  to the
Pontryagin density already at first order in $\alpha'$.
Following  Ref.~\cite{Cano:2021rey}, the  
generated dCS coupling is (denoted as $\alpha_\text{str}$) 
\begin{equation}
\alpha_{\rm str}
=
\frac{c}{2M_s^2},
\label{eq:alpha-string}
\end{equation}
where $M_s\equiv1/\sqrt{\alpha'}$ is the string scale. 
The explicit construction in \cite{Cano:2021rey} corresponds to $c=1$ (up to an overall sign). 
We retain $c$ to allow for possible order-one modifications associated
with other possible compactifications. 

Similar to the last UV completion of chiral fermions \ref{Fermionic}, using the chirality parameter in Eq.~(\ref{eq:Pi-alpha}), we can write it in terms of \eqref{eq:alpha-string} as
\beq
\PicS\approx
3.5\times 10^{-2}
\,c\,
\left(H\over M_s\right)^2
\sqrt{16\epsilon\over 0.032}.
\label{Pistringy}\eeq
Structurally, this is  similar to the earlier expression for $\Pic$ from the fermion UV completion in Eq.~(\ref{Pidet}). Here, we need $H\ll M_s$ for a consistent effective theory.

Again, inserting this into Eq.~(\ref{AoddrF}) gives
\beq
\Aodd_{\rm str} \approx 9.8\times 10^4\,c\,
\,X\,Y^{\rm str}\,Z
\eeq
where
\bea
&&Y^{\rm str}\equiv\left(H\over M_s\right)^2
\eea
and $X$ and $Z$ are given above in Eqs. \eqref{Xdef} and \eqref{Zdef}.
In this case, if we take $r$ near its upper bound of 0.032 (and comparable value of $16\epsilon$) giving $X\sim 1$, then it is technically possible to have $\Aodd_{\rm str}$ larger than the benchmark detectable value of $3.3\times 10^4$ from Eq.~(\ref{Adetect}). However, to do so, we need to be saturating the limit of the EFT. Namely, we need $H$ to be rather close to the string scale $M_s$. If we are deep into the controlled EFT regime of $H\ll M_s$, then $Y^{\rm str}\ll1$, and we have (for $X\sim 1$ and $H\ll M_s$)
\beq
|\Aodd_{\rm str}|\ll \AoddD.
\eeq
In fact, even for a very moderate separation  of $H\lesssim M_s/2$, we  already have $\Aodd_{\rm str}$ smaller than $\AoddD$.

Furthermore, if $r$ is much lower than its upper bound of 0.032, the situation can be improved, but at the expense of requiring extremely large $F$, as discussed above. Taking $\AoddD\approx 3.3\times 10^4$, the condition on $X$ for detectability ($|\Aodd_{\rm str}|\geq\AoddD$) is now 
\beq
c\,X\,Y^{\rm str}\,Z\geq 0.34.
\eeq
Using Eq.~(\ref{FXZform}) for $F$ and using this condition for detectability, we obtain
\beq
F\geq2.9\times10^7 c^{-1}X^3(Y^{\rm str})^{-1}
\eeq
Since $X\geq1$ by the existing bound on tensor modes and $Y^{\rm str}\ll1$ for a well behaved effective theory (which in turn requires $X\gg1$), we have the requirement 
\beq
F\gg 3\times 10^7\,\,\,\,(\mbox{strings in UV})
\eeq
for detectability. For the case of $F\sim N_e^4$, with $N_s\sim60$, this condition is close, but still not quite satisfied. 

\subsubsection*{KK Modes}

It is useful to express the string-scale suppression in terms of the
Kaluza--Klein scale,
\begin{equation}
\left(\frac{H}{M_s}\right)^2
=
\left(\frac{H}{M_{\rm KK}}\right)^2
\left(\frac{M_{\rm KK}}{M_s}\right)^2 .
\end{equation}
For a geometric compactification with radii at or above the string
length, $R\gtrsim\ell_s$, one has
\begin{equation}
M_{\rm KK}\sim\frac{1}{R}\lesssim
\frac{1}{\ell_s}=M_s.
\end{equation}
At the same time, a controlled four-dimensional effective description
during inflation requires the Kaluza--Klein modes to lie above the
Hubble scale \cite{Baumann:2014nda},
\begin{equation}
H<M_{\rm KK}.
\end{equation}
Thus, the hierarchy takes the form
\begin{equation}
H<M_{\rm KK}\lesssim M_s,
\end{equation}
and both factors obey $H/M_\text{KK}<1$ and $M_\text{KK}/M_s<1$, which
suppresses the parity-odd signal.  In a parametrically controlled
four-dimensional effective string theory, these ratios should provide
a separation between the inflationary, compactification, and string
scales.

The compactification in Ref.~\cite{Cano:2021rey} corresponds to the limiting
choice $R=\ell_s$, for which
\begin{equation}
M_{\rm KK}=M_s.
\end{equation}
More generally, increasing the
compactification radius lowers $M_{\rm KK}$ relative to $M_s$ and
introduces an additional hierarchy between the compactification and
string scales.

For completeness, the condition $M_{\rm KK}\lesssim M_s$ is specific
to the geometric regime $R\gtrsim\ell_s$.  Formally, taking
$R<\ell_s$ gives $M_{\rm KK}>M_s$, but this does not provide a
parametric improvement of the effective theory since winding states have
masses $M_w\sim R/\alpha'$,
and therefore become lighter than the string scale. Under $T$-duality (with $\alpha' = \ell_s^2$)
\begin{equation}
R\longleftrightarrow\frac{\alpha'}{R},
\end{equation}
relates  the small- and large-radius
descriptions.  Thus, restricting to the geometric regime
$R\gtrsim\ell_s$ entails no loss of a parametrically better hierarchy.
For $R<\ell_s$ the role of the light compactification tower is instead
played by winding states.



\subsection{Type IIB}

As another example, let us consider the type IIB theory with the Ramond-Ramond four-form gauge field $C_4$ and a stack of $N_7$ D7-branes. The relevant part of the bulk action is \cite{Polchinski:1998rr,Bachas:1999um,Giddings:2001yu}
\begin{align}
    S_\text{IIB} = &M_{10}^8\int_{\mathcal{M}_{10}}d^{10}X\sqrt{-G}\left( R_{10} - \frac{1}{4\cdot 5!}|F_5|^2 \right)\nonumber\\
    & + \frac{N_7 M_s^4}{3072\pi^5}
    \int_{W_8}C_4\wedge \mbox{Tr}(R\wedge R).\label{eq: D7 action setup b}
\end{align}
Here the 10-dimensional Planck mass is related to the string scale $M_s=1/\sqrt{\alpha'}$ by $M_{10}^8=M_s^8/(2\pi)^7$. Compactification on an Einstein frame volume $V_{6,E}$ gives the 4-dimensional Planck mass as
\beq
\mpl^2={M_s^8\over(2\pi)^7}V_{6,E}.
\eeq

We note that compactification leads to the general result \cite{Leedom:2024qgr}
\beq
\alpha={N_7\over192\pi^2 f\,\mpl}
\eeq
for the axion associated with a singular K\"ahler modulus. 
Here $f$ is the corresponding axion decay constant.
This is numerically related to our earlier result from integrating out fermions in Eq.~(\ref{fermionalpha}) if one replaces the axion decay constant by $f\to m/g$. However, in the present context the decay constant is related only to parameters of the compactification.

Here we consider the axion associated with the zero-mode of the Ramond-Ramond four-form $C_4$. After computing the decay constant, we find the following value for the dCS coupling
\beq
\alpha=c{N_7M_s^4 V_{4,E}\over 1536\pi^5\mpl^2},
\eeq
where $c$ is again an $\mathcal{O}(1)$ number depending on the compactification.
Then writing the Planck mass in terms of the string scale, we have
\beq
\alpha=c{\pi^2\over 12}{N_7\over M_s^4}{V_{4,E}\over V_{6,E}}.
\eeq

To gain a clearer understanding of the size of this, let us again consider the simple case of a $T^6=T^2\times T^2\times T^2$ toroidal compactification where $c=1$. This has
\beq
V_{4,E}=(2\pi R)^4,\,\,\,\,\,\,\,V_{6,E}=(2\pi R)^6.
\eeq
Hence, we have
\beq
\alpha={N_7\over 48M_s^4 R^2}.
\eeq
Now it is useful to note that in the Einstein frame the effective string scale $M_{s,\rm{eff}}$ is related to the original string scale $M_s$ and the string coupling $g_s$ by
\beq
M_{s,\rm{eff}}=g_s^{1/4}M_s.
\eeq
Using this we have
\beq
\alpha ={N_7\, g_s\over 48M_{s,\rm{eff}}^4 R^2}.
\eeq

The corresponding chirality parameter using Eq.~(\ref{eq:Pi-alpha}) is
\beq
\Pi_c\approx 1.5\times10^{-3}(N_7\,g_s)
\left(H\over M_{s,\rm{eff}}\right)^2
\left(M_{\rm KK}\over M_{s,\rm{eff}}\right)^2
\sqrt{16\epsilon\over 0.032},
\eeq
where we used $M_{\rm KK}=1/R$.
A controlled EFT requires $H\ll M_{s,\rm{eff}}$ and $M_{\rm KK}\lesssim M_{s,\rm{eff}}$. Taking $N_7$ large seems as though one could enhance this; however, perturbativity requires $N_7\,g_s\lesssim1$, so this still gives a small factor.\footnote{For open strings at the end of the D-branes, there exists Chan-Paton factors with $N_7$ labels that contribute $N_7^2$ combinations \cite{Polchinski:1998rq}. The species scale is then estimated to be $\Lambda_\text{sp}\sim \mpl/N_7$. However, in order for the Chan-Paton factors to be resolvable down to the string length, would require the species scale should not exceed the string scale, implying \cite{Dvali:2007hz,Dvali:2010vm} $N_7g_s\lesssim 1$.} 
This is an even slightly stronger suppression compared to Eq.~(\ref{Pistringy}).
Altogether this gives $\Pi_c\ll1$, and so the problems of the previous discussion follows. So one again needs  extreme enhancement of $F$.

\section{Conclusions}\label{Discussion}

In this paper, being motivated by the observability of parity violation in the gravitational sector during inflation generated by dCS, we placed observational and theoretical bounds on the corresponding coupling constant. Recognizing first that parity odd amplitude of the trispectrum is already bounded by the signal-to-noise ratio of a detectable signal, this enforced a numerical lower bound for future surveys given the number of modes and $k$-values. To go around this, theoretically we could imagine an enhancement factor $F$ to achieve detectability. Already with existing data, we are pushed to the edge of validity via the requirement $|\Pi_c|\leq 1$. But we showed that the UV completions impose a much tighter bound. 

From a UV completion of a chiral fermion, we saw that $\Pi_c$ itself was brought down to a value that much be much less than $5.9\times 10^{-3}$, further making the prospect of detectability worse. 
On string theoretic UV completions,  the heterotic compactification was only slightly higher.
Finally, we looked at type IIB string theory with a 4-form gauge field with a stack of $N_7$ D7-branes compactified on a $T^6$ torus. Naively, we saw that this could give an enhancement to the dCS coupling with an arbitrarily large $N_7$ number of D7-branes, but perturbativity required $N_7g_s\lesssim 1$, and consistency of the EFT again required  $\Pi_c$ to be very small. 
Altogether, any consistent EFT that allows for parity detectability  requires extreme enhancement of the tensor-scalar-scalar vertex. But, plenty of questions and avenues for future work with dCS remain. \footnote{It may be that the dCS Wilson coefficient under renormalization group flow vanishes in the IR at the end of inflation \cite{Baumann:2010ys,Assassi:2013gxa}. In this scenario, the inflaton is coupled to a conformal sector that is sequestered away from the end of inflation. This seems entirely possible especially if the $\beta$-equation for $\alpha$ arose from a UV completion in heterotic string theory \cite{Cano:2021rey,Alexander:2004xd}. In this manner, if the sequestered regime is within $k\eta\sim 10^3-10^4$, then dCS could still still produce leptogensis as well with $\Theta\lesssim 10^{-8}$ \cite{Alexander:2004us,Lyth:2005jf}, but there could be inflation constraints on type IIB similar to 
type IIA \cite{Hertzberg:2007wc} that have not been taken into account yet.}
In particular, to consider warping and various higher loop effects.\footnote{Finally, the full 3-point function between the inflaton and 2 gravitons only starts at 1-loop. While the loop may be suppressed, the mode function in the chiral fermion UV completion induces a chemical potential \cite{Qin:2026yuo,You:2026xoq,Bodas:2020yho,Bodas:2024hih,Bodas:2025vpb,Sundrum:2026ghf}, and thus Boltzmann suppression is reduced. This would be true for the scalar power spectrum, or if working with the effective dCS theory, true for the tensor power spectrum as well.}


\section*{Acknowledgments}
M.~P.~H. is supported in part by National Science Foundation grant PHY-2609905. A. P. C. would like to thank Soubhik Kumar for various discussions. 
This work used ChatGPT Astra to assist with literature review and numerics, all of which was verified and closely directed by the authors.


\bibliography{bilbio}

\end{document}

%% file: tikz_images.tex
\usetikzlibrary{calc,decorations.pathmorphing,decorations.markings,arrows.meta}
\tikzset{
  infl/.style={
    black,
    line cap=round,
    line width=.7pt
  },
  boundary/.style={
    gray!60,
    line cap=round,
    line width=2.0pt
  },
  vertex/.style={
    circle,
    fill=black,
    inner sep=1.1pt
  },
  source/.style={
    black,
    line width=.7pt,
    dash pattern=on 4pt off 3pt,
    line cap=round
  },
  sourceX/.style={
    black,
    line width=.85pt,
    line cap=round
  },
  energycut/.style={
    green!60!black,
    line width=2.2pt,
    line cap=round
  },
  gravitonwave/.style={
    black,
    line width=.7pt,
    decorate,
    decoration={snake, amplitude=1.15pt, segment length=6.5pt}
  }
}

\tikzset{
  fb/arrow/.style={
    postaction={decorate},
    decoration={markings,
      mark=at position 0.5 with {\arrow{Stealth[length=2.3mm,width=1.5mm]}}
    }
  },
  fb/scalar/.style={black,line width=.7pt,fb/arrow},
  fb/fermion/.style={black,line width=.7pt,
    dash pattern=on 4pt off 2.5pt,fb/arrow},
  fb/boundary/.style={gray!60,line width=2.0pt},
  fb/vertex/.style={circle,fill=black,inner sep=1.1pt}
}

\newcommand{\FermionBubbleCubicCore}[4]{%
\scalebox{#1}{%
\begin{tikzpicture}[
  line cap=round,
  line join=round,
  baseline=(fbBoundary.base),
  every node/.style={font=\large}
]
  \node[inner sep=0pt] (fbBoundary) at (0,2.0) {};
  \coordinate (fbL) at (-1.25,#2);
  \coordinate (fbR) at ( 1.25,#3);
  \coordinate (fbKL) at (-2.55,2.0);
  \coordinate (fbKR) at ( 2.55,2.0);

  \draw[fb/scalar] (fbL) -- (fbKL);
  \draw[fb/scalar] (fbKR) -- (fbR);

  \pgfmathsetmacro{\fbAngle}{atan2(#3-#2,2.5)}
  \pgfmathsetmacro{\fbRadius}{sqrt(2.5*2.5+(#3-#2)*(#3-#2))/2}
  \begin{scope}[shift={($(fbL)!0.5!(fbR)$)},rotate=\fbAngle]
    \draw[fb/fermion] (-\fbRadius,0)
      arc[start angle=180,end angle=0,
          x radius=\fbRadius cm,y radius=.72cm];
    \draw[fb/fermion] (\fbRadius,0)
      arc[start angle=0,end angle=-180,
          x radius=\fbRadius cm,y radius=.72cm];
    \ifnum#4=1\relax
      \node at (0,1.00) {$\psi$};
      \node at (0,-1.02) {$\overline{\psi}$};
    \fi
  \end{scope}

  \node[fb/vertex] at (fbL) {};
  \node[fb/vertex] at (fbR) {};
  \draw[fb/boundary] (-3.05,2.0) -- (3.05,2.0);
  \ifnum#4=1\relax
    \node[anchor=south] at (fbKL) {$\delta\varphi(\vec{k})$};
    \node[anchor=south] at (fbKR) {$\delta\varphi(-\vec{k})$};
  \fi
\end{tikzpicture}%
}%
}

\newcommand{\FermionBubbleQuarticCore}[3]{%
\scalebox{#1}{%
\begin{tikzpicture}[
  line cap=round,
  line join=round,
  baseline=(fbBoundary.base),
  every node/.style={font=\large}
]
  \node[inner sep=0pt] (fbBoundary) at (0,2.0) {};
  \coordinate (fbKL) at (-2.55,2.0);
  \coordinate (fbKR) at ( 2.55,2.0);
  \coordinate (fbV) at (0,{2.0-1.5*#2});
  \draw[fb/scalar] (fbV) -- (fbKL);
  \draw[fb/scalar] (fbKR) -- (fbV);

  \begin{scope}[shift={(fbV)},yscale=#2]
    \draw[fb/fermion] (0,0)
      arc[start angle=90,end angle=270,radius=.85cm];
    \draw[fb/fermion] (0,-1.70)
      arc[start angle=270,end angle=450,radius=.85cm];
    \ifnum#3=1\relax
      \node[anchor=east] at (-1.08,-.85) {$\psi$};
      \node[anchor=west] at ( 1.08,-.85) {$\overline{\psi}$};
    \fi
  \end{scope}

  \node[fb/vertex] at (fbV) {};
  \draw[fb/boundary] (-3.05,2.0) -- (3.05,2.0);
  \ifnum#3=1\relax
    \node[anchor=south] at (fbKL) {$\delta\varphi(\vec{k})$};
    \node[anchor=south] at (fbKR) {$\delta\varphi(-\vec{k})$};
  \fi
\end{tikzpicture}%
}%
}

%% file: trispectrum_momenta.tex
\usetikzlibrary{calc,arrows.meta,decorations.markings}

\newif\ifTMarrows
\newif\ifTMratios
\pgfkeys{
  /trispectrum/.is family,
  /trispectrum,
  q/.store in=\TMqinput,
  k3/.store in=\TMkthreeinput,
  theta1/.store in=\TMthetaoneinput,
  theta3/.store in=\TMthetathreeinput,
  chi/.store in=\TMchiinput,
  scale/.store in=\TMscale,
  length/.store in=\TMlength,
  rotate/.store in=\TMrotate,
  depth x/.store in=\TMdepthx,
  depth y/.store in=\TMdepthy,
  arrows/.is if=TMarrows,
  ratios/.is if=TMratios,
  layout/.is choice,
  layout/channel/.code={\def\TMlayout{0}},
  layout/rectangle/.code={\def\TMlayout{1}},
  defaults/.style={
    q=0.25,k3=1,theta1=pi/2,theta3=pi/2,chi=2*pi/3,
    scale=1,length=3.2,rotate=0,depth x=0.8,depth y=0.6,
    arrows=true,ratios=false,layout=channel
  }
}

\newcommand{\TrispectrumMomenta}[1][]{%
\begingroup
  \pgfkeys{/trispectrum,defaults,#1}%
  \pgfmathsetmacro{\TMq}{\TMqinput}%
  \pgfmathsetmacro{\TMkthree}{\TMkthreeinput}%
  \pgfmathsetmacro{\TMthetaone}{deg(\TMthetaoneinput)}%
  \pgfmathsetmacro{\TMthetathree}{deg(\TMthetathreeinput)}%
  \pgfmathsetmacro{\TMchi}{deg(\TMchiinput)}%
  \pgfmathtruncatemacro{\TMvalid}{(\TMq>0)&&(\TMkthree>0)}%
  \ifnum\TMvalid=0
    \PackageError{trispectrum-momenta}{q and k3 must be positive}%
      {Use a small positive q for the collapsed limit; the q direction is undefined at q=0.}%
  \else
    \pgfmathsetmacro{\TMxone}{sin(\TMthetaone)}%
    \pgfmathsetmacro{\TMzone}{cos(\TMthetaone)}%
    \pgfmathsetmacro{\TMxthree}{\TMkthree*sin(\TMthetathree)*cos(\TMchi)}%
    \pgfmathsetmacro{\TMythree}{\TMkthree*sin(\TMthetathree)*sin(\TMchi)}%
    \pgfmathsetmacro{\TMzthree}{\TMkthree*cos(\TMthetathree)}%
    \pgfmathsetmacro{\TMktwo}{sqrt(max(0,1+\TMq*\TMq-2*\TMq*\TMzone))}%
    \pgfmathsetmacro{\TMkfour}{sqrt(max(0,\TMkthree*\TMkthree+\TMq*\TMq+2*\TMq*\TMzthree))}%
    \pgfmathsetmacro{\TMsoftness}{\TMq/max(0.00001,min(1,\TMktwo,\TMkthree,\TMkfour))}%
    \begin{tikzpicture}[
      scale=\TMscale,transform shape,rotate=\TMrotate,
      x={(1cm,0cm)},y={(0cm,1cm)},z={(\TMdepthx cm,\TMdepthy cm)},
      line cap=round,line join=round,
      every node/.style={font=\small},
      TM/hard/.style={black,line width=0.95pt},
      TM/label/.style={fill=white,inner sep=1pt},
      TM/soft/.style={black,line width=0.6pt,dash pattern=on 1pt off 0.8pt}
    ]
      \ifTMarrows
        \tikzset{TM/hard/.append style={postaction={decorate},
          decoration={markings,mark=at position 0.55 with
            {\arrow{Stealth[length=4pt,width=3pt]}}}}}%
      \fi
      \def\TMdrawleg##1##2##3{%
        \draw[TM/hard] (##1)--(##2);
        \coordinate (TM-edgemid) at ($(##1)!0.5!(##2)$);
        \node[TM/label] at ($(TM-edgemid)!8pt!-90:(##2)$) {$##3$};
      }%
      \coordinate (TM-O) at (0,0,0);
      \coordinate (TM-A) at ({\TMlength*\TMxone},0,{\TMlength*\TMzone});
      \coordinate (TM-C) at ({\TMlength*\TMxthree},{\TMlength*\TMythree},
        {\TMlength*(\TMq+\TMzthree)});
      \ifnum\TMlayout=0
        \coordinate (TM-B) at (0,0,{\TMlength*\TMq});
        \coordinate (TM-Qstart) at (TM-O);
        \coordinate (TM-Qend) at (TM-B);
        \TMdrawleg{TM-O}{TM-A}{k_1}
        \TMdrawleg{TM-A}{TM-B}{k_2}
        \TMdrawleg{TM-B}{TM-C}{k_3}
        \TMdrawleg{TM-C}{TM-O}{k_4}
      \else
        \coordinate (TM-B) at ({\TMlength*(\TMxone+\TMxthree)},
          {\TMlength*\TMythree},{\TMlength*(\TMzone+\TMzthree)});
        \coordinate (TM-Qstart) at ({0.5*\TMlength*(\TMxone+\TMxthree)},
          {0.5*\TMlength*\TMythree},
          {\TMlength*(0.5*(\TMzone+\TMzthree)-0.25*\TMq)});
        \coordinate (TM-Qend) at ({0.5*\TMlength*(\TMxone+\TMxthree)},
          {0.5*\TMlength*\TMythree},
          {\TMlength*(0.5*(\TMzone+\TMzthree)+0.75*\TMq)});
        \TMdrawleg{TM-O}{TM-A}{k_1}
        \TMdrawleg{TM-A}{TM-B}{k_3}
        \TMdrawleg{TM-B}{TM-C}{k_2}
        \TMdrawleg{TM-C}{TM-O}{k_4}
      \fi
      \draw[TM/soft] (TM-Qstart)--(TM-Qend)
        node[midway,above=3pt,fill=white,inner sep=0.8pt] {$q$};
      \foreach \TMpoint in {TM-O,TM-A,TM-B,TM-C}
        \fill (\TMpoint) circle (1.15pt);
      \ifTMratios
        \path (current bounding box.south) ++(0,-0.24)
          node[anchor=north,align=center,font=\scriptsize] {
          $q/k_1=\pgfmathprintnumber[fixed,precision=2]{\TMq},\quad
           k_2/k_1=\pgfmathprintnumber[fixed,precision=2]{\TMktwo}$\\[-1pt]
          $k_3/k_1=\pgfmathprintnumber[fixed,precision=2]{\TMkthree},\quad
           k_4/k_1=\pgfmathprintnumber[fixed,precision=2]{\TMkfour}$\\[-1pt]
          $q/\min_i k_i=\pgfmathprintnumber[fixed,precision=2]{\TMsoftness}$};
      \fi
    \end{tikzpicture}%
  \fi
\endgroup
}


%% file: bilbio.bib
@article{Starobinsky:1980te,
    author = "Starobinsky, Alexei A.",
    editor = "Khalatnikov, I. M. and Mineev, V. P.",
    title = "{A New Type of Isotropic Cosmological Models Without Singularity}",
    doi = "10.1016/0370-2693(80)90670-X",
    journal = "Phys. Lett. B",
    volume = "91",
    pages = "99--102",
    year = "1980"
}

@article{Guth:1980zm,
    author = "Guth, Alan H.",
    editor = "Fang, Li-Zhi and Ruffini, R.",
    title = "{The Inflationary Universe: A Possible Solution to the Horizon and Flatness Problems}",
    reportNumber = "SLAC-PUB-2576",
    doi = "10.1103/PhysRevD.23.347",
    journal = "Phys. Rev. D",
    volume = "23",
    pages = "347--356",
    year = "1981"
}

@article{Linde:1981mu,
    author = "Linde, Andrei D.",
    editor = "Fang, Li-Zhi and Ruffini, R.",
    title = "{A New Inflationary Universe Scenario: A Possible Solution of the Horizon, Flatness, Homogeneity, Isotropy and Primordial Monopole Problems}",
    reportNumber = "LEBEDEV-81-229",
    doi = "10.1016/0370-2693(82)91219-9",
    journal = "Phys. Lett. B",
    volume = "108",
    pages = "389--393",
    year = "1982"
}

@article{Albrecht:1982wi,
    author = "Albrecht, Andreas and Steinhardt, Paul J.",
    editor = "Fang, Li-Zhi and Ruffini, R.",
    title = "{Cosmology for Grand Unified Theories with Radiatively Induced Symmetry Breaking}",
    reportNumber = "UPR-0185T",
    doi = "10.1103/PhysRevLett.48.1220",
    journal = "Phys. Rev. Lett.",
    volume = "48",
    pages = "1220--1223",
    year = "1982"
}

@article{Linde:1983gd,
    author = "Linde, Andrei D.",
    title = "{Chaotic Inflation}",
    doi = "10.1016/0370-2693(83)90837-7",
    journal = "Phys. Lett. B",
    volume = "129",
    pages = "177--181",
    year = "1983"
}

@article{Mukhanov:1981xt,
    author = "Mukhanov, Viatcheslav F. and Chibisov, G. V.",
    title = "{Quantum Fluctuations and a Nonsingular Universe}",
    journal = "JETP Lett.",
    volume = "33",
    pages = "532--535",
    year = "1981"
}

@article{Hawking:1982cz,
    author = "Hawking, S. W.",
    title = "{The Development of Irregularities in a Single Bubble Inflationary Universe}",
    reportNumber = "Print-83-0015 (CAMBRIDGE)",
    doi = "10.1016/0370-2693(82)90373-2",
    journal = "Phys. Lett. B",
    volume = "115",
    pages = "295",
    year = "1982"
}

@article{Starobinsky:1982ee,
    author = "Starobinsky, Alexei A.",
    title = "{Dynamics of Phase Transition in the New Inflationary Universe Scenario and Generation of Perturbations}",
    doi = "10.1016/0370-2693(82)90541-X",
    journal = "Phys. Lett. B",
    volume = "117",
    pages = "175--178",
    year = "1982"
}

@article{Guth:1982ec,
    author = "Guth, Alan H. and Pi, S. Y.",
    title = "{Fluctuations in the New Inflationary Universe}",
    doi = "10.1103/PhysRevLett.49.1110",
    journal = "Phys. Rev. Lett.",
    volume = "49",
    pages = "1110--1113",
    year = "1982"
}

@article{PhysRevD.28.679,
  title = {Spontaneous creation of almost scale-free density perturbations in an inflationary universe},
  author = {Bardeen, James M. and Steinhardt, Paul J. and Turner, Michael S.},
  journal = {Phys. Rev. D},
  volume = {28},
  issue = {4},
  pages = {679--693},
  numpages = {0},
  year = {1983},
  month = {Aug},
  publisher = {American Physical Society},
  doi = {10.1103/PhysRevD.28.679},
  url = {https://link.aps.org/doi/10.1103/PhysRevD.28.679}
}

@article{Creque-Sarbinowski:2023wmb,
    author = "Creque-Sarbinowski, Cyril and Alexander, Stephon and Kamionkowski, Marc and Philcox, Oliver",
    title = "{Parity-violating trispectrum from Chern-Simons gravity}",
    eprint = "2303.04815",
    archivePrefix = "arXiv",
    primaryClass = "astro-ph.CO",
    doi = "10.1088/1475-7516/2023/11/029",
    journal = "JCAP",
    volume = "11",
    pages = "029",
    year = "2023"
}

@article{Cassem:2026aib,
    author = "Cassem, Alexander and Hertzberg, Mark P.",
    title = "{Theoretical and Observational Bounds on Dynamical Chern-Simons Gravity as an Effective Field Theory}",
    eprint = "2604.07332",
    archivePrefix = "arXiv",
    primaryClass = "hep-th",
    month = "4",
    year = "2026"
}

@article{Philcox:2024wqx,
    author = "Philcox, Oliver H. E. and Shiraishi, Maresuke",
    title = "{Non-Gaussianity beyond the scalar sector: A search for tensor and mixed tensor-scalar bispectra with Planck data}",
    eprint = "2409.10595",
    archivePrefix = "arXiv",
    primaryClass = "astro-ph.CO",
    doi = "10.1103/pb7x-lp19",
    journal = "Phys. Rev. D",
    volume = "111",
    number = "12",
    pages = "123502",
    year = "2025"
}

@article{Cano:2021rey,
    author = "Cano, Pablo A. and Ruip{\'e}rez, Alejandro",
    title = "{String gravity in D=4}",
    eprint = "2111.04750",
    archivePrefix = "arXiv",
    primaryClass = "hep-th",
    doi = "10.1103/PhysRevD.105.044022",
    journal = "Phys. Rev. D",
    volume = "105",
    number = "4",
    pages = "044022",
    year = "2022"
}

@article{Baumann:2010ys,
    author = "Baumann, Daniel and Green, Daniel",
    title = "{Desensitizing Inflation from the Planck Scale}",
    eprint = "1004.3801",
    archivePrefix = "arXiv",
    primaryClass = "hep-th",
    doi = "10.1007/JHEP09(2010)057",
    journal = "JHEP",
    volume = "09",
    pages = "057",
    year = "2010"
}

@article{Assassi:2013gxa,
    author = "Assassi, Valentin and Baumann, Daniel and Green, Daniel and McAllister, Liam",
    title = "{Planck-Suppressed Operators}",
    eprint = "1304.5226",
    archivePrefix = "arXiv",
    primaryClass = "hep-th",
    doi = "10.1088/1475-7516/2014/01/033",
    journal = "JCAP",
    volume = "01",
    pages = "033",
    year = "2014"
}

@article{Alexander:2004xd,
    author = "Alexander, Stephon H. S. and Gates, Jr., S. James",
    title = "{Can the string scale be related to the cosmic baryon asymmetry?}",
    eprint = "hep-th/0409014",
    archivePrefix = "arXiv",
    reportNumber = "SLAC-PUB-10688, UMDEPP-05-014",
    doi = "10.1088/1475-7516/2006/06/018",
    journal = "JCAP",
    volume = "06",
    pages = "018",
    year = "2006"
}

@article{Alexander:2004us,
    author = "Alexander, Stephon Haigh-Solom and Peskin, Michael E. and Sheikh-Jabbari, Mohammad M.",
    title = "{Leptogenesis from gravity waves in models of inflation}",
    eprint = "hep-th/0403069",
    archivePrefix = "arXiv",
    reportNumber = "SLAC-PUB-10226, SU-ITP-04-08",
    doi = "10.1103/PhysRevLett.96.081301",
    journal = "Phys. Rev. Lett.",
    volume = "96",
    pages = "081301",
    year = "2006"
}

@article{Lyth:2005jf,
    author = "Lyth, David H. and Quimbay, Carlos and Rodriguez, Yeinzon",
    title = "{Leptogenesis and tensor polarisation from a gravitational Chern-Simons term}",
    eprint = "hep-th/0501153",
    archivePrefix = "arXiv",
    doi = "10.1088/1126-6708/2005/03/016",
    journal = "JHEP",
    volume = "03",
    pages = "016",
    year = "2005"
}

@article{Alexander:2004wk,
    author = "Alexander, Stephon and Martin, Jerome",
    title = "{Birefringent gravitational waves and the consistency check of inflation}",
    eprint = "hep-th/0410230",
    archivePrefix = "arXiv",
    reportNumber = "SLAC-PUB-10816",
    doi = "10.1103/PhysRevD.71.063526",
    journal = "Phys. Rev. D",
    volume = "71",
    pages = "063526",
    year = "2005"
}

@article{Alexander:2009tp,
    author = "Alexander, Stephon and Yunes, Nicolas",
    title = "{Chern-Simons Modified General Relativity}",
    eprint = "0907.2562",
    archivePrefix = "arXiv",
    primaryClass = "hep-th",
    doi = "10.1016/j.physrep.2009.07.002",
    journal = "Phys. Rept.",
    volume = "480",
    pages = "1--55",
    year = "2009"
}

@article{Satoh:2008ck,
    author = "Satoh, Masaki and Soda, Jiro",
    title = "{Higher Curvature Corrections to Primordial Fluctuations in Slow-roll Inflation}",
    eprint = "0806.4594",
    archivePrefix = "arXiv",
    primaryClass = "astro-ph",
    doi = "10.1088/1475-7516/2008/09/019",
    journal = "JCAP",
    volume = "09",
    pages = "019",
    year = "2008"
}

@article{Green:1984sg,
    author = "Green, Michael B. and Schwarz, John H.",
    title = "{Anomaly Cancellation in Supersymmetric D=10 Gauge Theory and Superstring Theory}",
    reportNumber = "CALT-68-1182",
    doi = "10.1016/0370-2693(84)91565-X",
    journal = "Phys. Lett. B",
    volume = "149",
    pages = "117--122",
    year = "1984"
}

@book{Polchinski:1998rr,
    author = "Polchinski, J.",
    title = "{String theory. Vol. 2: Superstring theory and beyond}",
    doi = "10.1017/CBO9780511618123",
    isbn = "978-0-511-25228-0, 978-0-521-63304-8, 978-0-521-67228-3",
    publisher = "Cambridge University Press",
    series = "Cambridge Monographs on Mathematical Physics",
    month = "12",
    year = "2007"
}

@article{Planck:2018vyg,
    author = "Aghanim, N. and others",
    collaboration = "Planck",
    title = "{Planck 2018 results. VI. Cosmological parameters}",
    eprint = "1807.06209",
    archivePrefix = "arXiv",
    primaryClass = "astro-ph.CO",
    doi = "10.1051/0004-6361/201833910",
    journal = "Astron. Astrophys.",
    volume = "641",
    pages = "A6",
    year = "2020",
    note = "[Erratum: Astron.Astrophys. 652, C4 (2021)]"
}

@article{Planck:2018jri,
    author = "Akrami, Y. and others",
    collaboration = "Planck",
    title = "{Planck 2018 results. X. Constraints on inflation}",
    eprint = "1807.06211",
    archivePrefix = "arXiv",
    primaryClass = "astro-ph.CO",
    doi = "10.1051/0004-6361/201833887",
    journal = "Astron. Astrophys.",
    volume = "641",
    pages = "A10",
    year = "2020"
}

@article{Coulton:2023oug,
    author = "Coulton, William R. and Philcox, Oliver H. E. and Villaescusa-Navarro, Francisco",
    title = "{Signatures of a parity-violating universe}",
    eprint = "2306.11782",
    archivePrefix = "arXiv",
    primaryClass = "astro-ph.CO",
    doi = "10.1103/PhysRevD.109.023531",
    journal = "Phys. Rev. D",
    volume = "109",
    number = "2",
    pages = "023531",
    year = "2024"
}

@article{Diego-Palazuelos:2022cnh,
    author = "Diego-Palazuelos, P. and others",
    title = "{Robustness of cosmic birefringence measurement against Galactic foreground emission and instrumental systematics}",
    eprint = "2210.07655",
    archivePrefix = "arXiv",
    primaryClass = "astro-ph.CO",
    doi = "10.1088/1475-7516/2023/01/044",
    journal = "JCAP",
    volume = "01",
    pages = "044",
    year = "2023"
}

@article{Hou:2022wfj,
    author = "Hou, Jiamin and Slepian, Zachary and Cahn, Robert N.",
    title = "{Measurement of parity-odd modes in the large-scale 4-point correlation function of Sloan Digital Sky Survey Baryon Oscillation Spectroscopic Survey twelfth data release CMASS and LOWZ galaxies}",
    eprint = "2206.03625",
    archivePrefix = "arXiv",
    primaryClass = "astro-ph.CO",
    doi = "10.1093/mnras/stad1062",
    journal = "Mon. Not. Roy. Astron. Soc.",
    volume = "522",
    number = "4",
    pages = "5701--5739",
    year = "2023"
}

@article{Minami:2020odp,
    author = "Minami, Yuto and Komatsu, Eiichiro",
    title = "{New Extraction of the Cosmic Birefringence from the Planck 2018 Polarization Data}",
    eprint = "2011.11254",
    archivePrefix = "arXiv",
    primaryClass = "astro-ph.CO",
    doi = "10.1103/PhysRevLett.125.221301",
    journal = "Phys. Rev. Lett.",
    volume = "125",
    number = "22",
    pages = "221301",
    year = "2020"
}

@article{Philcox:2022hkh,
    author = "Philcox, Oliver H. E.",
    title = "{Probing parity violation with the four-point correlation function of BOSS galaxies}",
    eprint = "2206.04227",
    archivePrefix = "arXiv",
    primaryClass = "astro-ph.CO",
    doi = "10.1103/PhysRevD.106.063501",
    journal = "Phys. Rev. D",
    volume = "106",
    number = "6",
    pages = "063501",
    year = "2022"
}

@article{Krolewski:2024paz,
    author = "Krolewski, Alex and May, Simon and Smith, Kendrick and Hopkins, Hans",
    title = "{No evidence for parity violation in BOSS}",
    eprint = "2407.03397",
    archivePrefix = "arXiv",
    primaryClass = "astro-ph.CO",
    doi = "10.1088/1475-7516/2024/08/044",
    journal = "JCAP",
    volume = "08",
    pages = "044",
    year = "2024"
}

@article{Tristram:2021tvh,
    author = "Tristram, M. and others",
    title = "{Improved limits on the tensor-to-scalar ratio using BICEP and Planck data}",
    eprint = "2112.07961",
    archivePrefix = "arXiv",
    primaryClass = "astro-ph.CO",
    doi = "10.1103/PhysRevD.105.083524",
    journal = "Phys. Rev. D",
    volume = "105",
    number = "8",
    pages = "083524",
    year = "2022"
}

@article{Sailer:2021yzm,
    author = "Sailer, Noah and Castorina, Emanuele and Ferraro, Simone and White, Martin",
    title = "{Cosmology at high redshift {\textemdash} a probe of fundamental physics}",
    eprint = "2106.09713",
    archivePrefix = "arXiv",
    primaryClass = "astro-ph.CO",
    doi = "10.1088/1475-7516/2021/12/049",
    journal = "JCAP",
    volume = "12",
    number = "12",
    pages = "049",
    year = "2021"
}

@article{Cahn:2021ltp,
    author = "Cahn, Robert N. and Slepian, Zachary and Hou, Jiamin",
    title = "{Test for Cosmological Parity Violation Using the 3D Distribution of Galaxies}",
    eprint = "2110.12004",
    archivePrefix = "arXiv",
    primaryClass = "astro-ph.CO",
    doi = "10.1103/PhysRevLett.130.201002",
    journal = "Phys. Rev. Lett.",
    volume = "130",
    number = "20",
    pages = "201002",
    year = "2023"
}

@inbook{Alvarez-Gaume:2022aak,
    author = "Alvarez-Gaume, Luis and Vazquez-Mozo, Miguel A.",
    title = "{Anomalies and the Green-Schwarz Mechanism}",
    eprint = "2211.06467",
    archivePrefix = "arXiv",
    primaryClass = "hep-th",
    doi = "10.1007/978-981-19-3079-9_68-1",
    year = "2024"
}

@article{Svrcek:2006yi,
    author = "Svrcek, Peter and Witten, Edward",
    title = "{Axions In String Theory}",
    eprint = "hep-th/0605206",
    archivePrefix = "arXiv",
    reportNumber = "SLAC-PUB-11894",
    doi = "10.1088/1126-6708/2006/06/051",
    journal = "JHEP",
    volume = "06",
    pages = "051",
    year = "2006"
}

@article{Stefanski:1998he,
    author = "Stefanski, Jr., Bogdan",
    title = "{Gravitational couplings of D-branes and O-planes}",
    eprint = "hep-th/9812088",
    archivePrefix = "arXiv",
    reportNumber = "DAMTP-1998-166",
    doi = "10.1016/S0550-3213(99)00147-9",
    journal = "Nucl. Phys. B",
    volume = "548",
    pages = "275--290",
    year = "1999"
}

@article{Weissenbacher:2020cyf,
    author = "Weissenbacher, Matthias",
    title = "{On $\alpha'$-effects from $D$-branes in $4d \; \mathcal{N} = 1$}",
    eprint = "2006.15552",
    archivePrefix = "arXiv",
    primaryClass = "hep-th",
    doi = "10.1007/JHEP11(2020)076",
    journal = "JHEP",
    volume = "11",
    pages = "076",
    year = "2020"
}

@article{Bartolo:2017szm,
    author = "Bartolo, Nicola and Orlando, Giorgio",
    title = "{Parity breaking signatures from a Chern-Simons coupling during inflation: the case of non-Gaussian gravitational waves}",
    eprint = "1706.04627",
    archivePrefix = "arXiv",
    primaryClass = "astro-ph.CO",
    doi = "10.1088/1475-7516/2017/07/034",
    journal = "JCAP",
    volume = "07",
    pages = "034",
    year = "2017"
}

@article{Seery:2008ax,
    author = "Seery, David and Sloth, Martin S. and Vernizzi, Filippo",
    title = "{Inflationary trispectrum from graviton exchange}",
    eprint = "0811.3934",
    archivePrefix = "arXiv",
    primaryClass = "astro-ph",
    doi = "10.1088/1475-7516/2009/03/018",
    journal = "JCAP",
    volume = "03",
    pages = "018",
    year = "2009"
}

@article{Maldacena:2002vr,
    author = "Maldacena, Juan Martin",
    title = "{Non-Gaussian features of primordial fluctuations in single field inflationary models}",
    eprint = "astro-ph/0210603",
    archivePrefix = "arXiv",
    doi = "10.1088/1126-6708/2003/05/013",
    journal = "JHEP",
    volume = "05",
    pages = "013",
    year = "2003"
}

@article{Cabass:2022rhr,
    author = "Cabass, Giovanni and Jazayeri, Sadra and Pajer, Enrico and Stefanyszyn, David",
    title = "{Parity violation in the scalar trispectrum: no-go theorems and yes-go examples}",
    eprint = "2210.02907",
    archivePrefix = "arXiv",
    primaryClass = "hep-th",
    doi = "10.1007/JHEP02(2023)021",
    journal = "JHEP",
    volume = "02",
    pages = "021",
    year = "2023"
}

@article{Smith:2015uia,
    author = "Smith, Kendrick M. and Senatore, Leonardo and Zaldarriaga, Matias",
    title = "{Optimal analysis of the CMB trispectrum}",
    eprint = "1502.00635",
    archivePrefix = "arXiv",
    primaryClass = "astro-ph.CO",
    month = "2",
    year = "2015"
}

@article{Philcox:2025bvj,
    author = "Philcox, Oliver H. E.",
    title = "{Searching for inflationary physics with the CMB trispectrum. I. Primordial theory and optimal estimators}",
    eprint = "2502.04434",
    archivePrefix = "arXiv",
    primaryClass = "astro-ph.CO",
    doi = "10.1103/wkj2-q62t",
    journal = "Phys. Rev. D",
    volume = "111",
    number = "12",
    pages = "123532",
    year = "2025"
}

@article{Philcox:2025lrr,
    author = "Philcox, Oliver H. E.",
    title = "{Searching for inflationary physics with the CMB trispectrum. II. Code and validation}",
    eprint = "2502.05258",
    archivePrefix = "arXiv",
    primaryClass = "astro-ph.CO",
    doi = "10.1103/bprw-3zj1",
    journal = "Phys. Rev. D",
    volume = "111",
    number = "12",
    pages = "123533",
    year = "2025"
}

@article{Philcox:2025wts,
    author = "Philcox, Oliver H. E.",
    title = "{Searching for inflationary physics with the CMB trispectrum. III. Constraints from Planck}",
    eprint = "2502.06931",
    archivePrefix = "arXiv",
    primaryClass = "astro-ph.CO",
    doi = "10.1103/y81z-g7th",
    journal = "Phys. Rev. D",
    volume = "111",
    number = "12",
    pages = "123534",
    year = "2025"
}

@article{Senatore:2009gt,
    author = "Senatore, Leonardo and Smith, Kendrick M. and Zaldarriaga, Matias",
    title = "{Non-Gaussianities in Single Field Inflation and their Optimal Limits from the WMAP 5-year Data}",
    eprint = "0905.3746",
    archivePrefix = "arXiv",
    primaryClass = "astro-ph.CO",
    doi = "10.1088/1475-7516/2010/01/028",
    journal = "JCAP",
    volume = "01",
    pages = "028",
    year = "2010"
}

@article{Gordon:2000hv,
    author = "Gordon, Christopher and Wands, David and Bassett, Bruce A. and Maartens, Roy",
    title = "{Adiabatic and entropy perturbations from inflation}",
    eprint = "astro-ph/0009131",
    archivePrefix = "arXiv",
    doi = "10.1103/PhysRevD.63.023506",
    journal = "Phys. Rev. D",
    volume = "63",
    pages = "023506",
    year = "2000"
}

@article{Toms:2018wpy,
    author = "Toms, David J.",
    title = "{Effective action for the Yukawa model in curved spacetime}",
    eprint = "1804.08350",
    archivePrefix = "arXiv",
    primaryClass = "hep-th",
    doi = "10.1007/JHEP05(2018)139",
    journal = "JHEP",
    volume = "05",
    pages = "139",
    year = "2018"
}

@article{Alexander:2022cow,
    author = "Alexander, Stephon and Creque-Sarbinowski, Cyril",
    title = "{UV completions of Chern-Simons gravity}",
    eprint = "2207.05094",
    archivePrefix = "arXiv",
    primaryClass = "hep-ph",
    doi = "10.1103/PhysRevD.108.104046",
    journal = "Phys. Rev. D",
    volume = "108",
    number = "10",
    pages = "104046",
    year = "2023"
}

@article{Alexander:2024vav,
    author = "Alexander, Stephon and Bernardo, Heliudson and Creque-Sarbinowski, Cyril",
    title = "{Nontriviality of dynamical Chern-Simons gravity and the standard model}",
    eprint = "2403.15657",
    archivePrefix = "arXiv",
    primaryClass = "hep-th",
    doi = "10.1103/PhysRevD.110.025015",
    journal = "Phys. Rev. D",
    volume = "110",
    number = "2",
    pages = "025015",
    year = "2024"
}

@article{Dvali:2007hz,
    author = "Dvali, Gia",
    title = "{Black Holes and Large N Species Solution to the Hierarchy Problem}",
    eprint = "0706.2050",
    archivePrefix = "arXiv",
    primaryClass = "hep-th",
    doi = "10.1002/prop.201000009",
    journal = "Fortsch. Phys.",
    volume = "58",
    pages = "528--536",
    year = "2010"
}

@article{Gross:1986mw,
    author = "Gross, David J. and Sloan, John H.",
    title = "{The Quartic Effective Action for the Heterotic String}",
    reportNumber = "NSF-ITP-87-02",
    doi = "10.1016/0550-3213(87)90465-2",
    journal = "Nucl. Phys. B",
    volume = "291",
    pages = "41--89",
    year = "1987"
}

@book{Baumann:2014nda,
    author = "Baumann, Daniel and McAllister, Liam",
    title = "{Inflation and String Theory}",
    eprint = "1404.2601",
    archivePrefix = "arXiv",
    primaryClass = "hep-th",
    doi = "10.1017/CBO9781316105733",
    isbn = "978-1-107-08969-3, 978-1-316-23718-2",
    publisher = "Cambridge University Press",
    series = "Cambridge Monographs on Mathematical Physics",
    month = "5",
    year = "2015"
}

@article{Qin:2026yuo,
    author = "Qin, Zhehan and Xianyu, Zhong-Zhi",
    title = "{Cosmological Collider Signals From a Triangle Loop}",
    eprint = "2609.09288",
    archivePrefix = "arXiv",
    primaryClass = "hep-th",
    month = "9",
    year = "2026"
}

@article{You:2026xoq,
    author = "You, Jingtao and Song, Linghao and Han, Chengcheng and He, Hong-Jian and Chen, Xingang and Xianyu, Zhong-Zhi",
    title = "{Cosmological Collider Signatures from Right-Handed Neutrino Loop}",
    eprint = "2605.21419",
    archivePrefix = "arXiv",
    primaryClass = "hep-ph",
    month = "5",
    year = "2026"
}

@article{Bodas:2020yho,
    author = "Bodas, Arushi and Kumar, Soubhik and Sundrum, Raman",
    title = "{The Scalar Chemical Potential in Cosmological Collider Physics}",
    eprint = "2010.04727",
    archivePrefix = "arXiv",
    primaryClass = "hep-ph",
    reportNumber = "UMD-PP-020-09",
    doi = "10.1007/JHEP02(2021)079",
    journal = "JHEP",
    volume = "02",
    pages = "079",
    year = "2021"
}

@article{Sundrum:2026ghf,
    author = "Sundrum, Raman and Xu, Zhaohui",
    title = "{Extra-dimensional Origins of Chemical Potentials at the Cosmological Collider}",
    eprint = "2608.00158",
    archivePrefix = "arXiv",
    primaryClass = "hep-ph",
    month = "7",
    year = "2026"
}

@article{Bodas:2025vpb,
    author = "Bodas, Arushi and Broadberry, Edward and Sundrum, Raman and Xu, Zhaohui",
    title = "{Charged loops at the cosmological collider with chemical potential}",
    eprint = "2507.22978",
    archivePrefix = "arXiv",
    primaryClass = "hep-ph",
    reportNumber = "FERMILAB-PUB-25-0519-V",
    doi = "10.1007/JHEP01(2026)083",
    journal = "JHEP",
    volume = "01",
    pages = "083",
    year = "2026"
}

@article{Bodas:2024hih,
    author = "Bodas, Arushi and Broadberry, Edward and Sundrum, Raman",
    title = "{Grand unification at the cosmological collider with chemical potential}",
    eprint = "2409.07524",
    archivePrefix = "arXiv",
    primaryClass = "hep-ph",
    reportNumber = "FERMILAB-PUB-24-0568-V",
    doi = "10.1007/JHEP01(2025)115",
    journal = "JHEP",
    volume = "01",
    pages = "115",
    year = "2025"
}

@inproceedings{Baumann:2009ds,
    author = "Baumann, Daniel",
    title = "{Inflation}",
    booktitle = "{Theoretical Advanced Study Institute in Elementary Particle Physics}: {Physics of the Large and the Small}",
    eprint = "0907.5424",
    archivePrefix = "arXiv",
    primaryClass = "hep-th",
    reportNumber = "TASI-2009",
    doi = "10.1142/9789814327183_0010",
    pages = "523--686",
    year = "2011"
}

@article{Hertzberg:2007wc,
    author = "Hertzberg, Mark P. and Kachru, Shamit and Taylor, Washington and Tegmark, Max",
    title = "{Inflationary Constraints on Type IIA String Theory}",
    eprint = "0711.2512",
    archivePrefix = "arXiv",
    primaryClass = "hep-th",
    reportNumber = "MIT-CTP-3905, SLAC-PUB-12999",
    doi = "10.1088/1126-6708/2007/12/095",
    journal = "JHEP",
    volume = "12",
    pages = "095",
    year = "2007"
}

@article{Giddings:2001yu,
    author = "Giddings, Steven B. and Kachru, Shamit and Polchinski, Joseph",
    title = "{Hierarchies from fluxes in string compactifications}",
    eprint = "hep-th/0105097",
    archivePrefix = "arXiv",
    reportNumber = "SLAC-PUB-8807, NSF-ITP-01-37, SU-ITP-01-16",
    doi = "10.1103/PhysRevD.66.106006",
    journal = "Phys. Rev. D",
    volume = "66",
    pages = "106006",
    year = "2002"
}

@article{Leedom:2024qgr,
    author = "Leedom, Jacob M. and Putti, Margherita and Righi, Nicole and Westphal, Alexander",
    title = "{Preheating axions in string cosmology}",
    eprint = "2411.18496",
    archivePrefix = "arXiv",
    primaryClass = "hep-th",
    reportNumber = "DESY-24-163, KCL-PH-TH/2024-60",
    doi = "10.1007/JHEP04(2025)095",
    journal = "JHEP",
    volume = "04",
    pages = "095",
    year = "2025"
}

@article{Bachas:1999um,
    author = "Bachas, Constantin P. and Bain, Pascal and Green, Michael B.",
    title = "{Curvature terms in D-brane actions and their M theory origin}",
    eprint = "hep-th/9903210",
    archivePrefix = "arXiv",
    reportNumber = "LPTENS-99-07, DAMTP-1999-14",
    doi = "10.1088/1126-6708/1999/05/011",
    journal = "JHEP",
    volume = "05",
    pages = "011",
    year = "1999"
}

@article{Dvali:2010vm,
    author = "Dvali, Gia and Gomez, Cesar",
    title = "{Species and Strings}",
    eprint = "1004.3744",
    archivePrefix = "arXiv",
    primaryClass = "hep-th",
    reportNumber = "CERN-PH-TH-2010-069, IFT-UAM-CSIC-10-25",
    month = "4",
    year = "2010"
}

@book{Polchinski:1998rq,
    author = "Polchinski, J.",
    title = "{String theory. Vol. 1: An introduction to the bosonic string}",
    doi = "10.1017/CBO9780511816079",
    isbn = "978-0-511-25227-3, 978-0-521-67227-6, 978-0-521-63303-1",
    publisher = "Cambridge University Press",
    series = "Cambridge Monographs on Mathematical Physics",
    month = "12",
    year = "2007"
}
